\documentclass[a4paper,fleqn]{cas-dc}

\usepackage[numbers]{natbib}

\def\tsc#1{\csdef{#1}{\textsc{\lowercase{#1}}\xspace}}
\tsc{WGM}
\tsc{QE}

\begin{document}
\let\WriteBookmarks\relax
\def\floatpagepagefraction{1}
\def\textpagefraction{.001}

\shorttitle{<short title of the paper for running head>}    

\shortauthors{<short author list for running head>}  

\title [mode = title]{Lesion-Aware Cross-Phase Attention Network for Renal Tumor Subtype Classification on Multi-Phase CT Scans}  



%

\author[1]{Kwang-Hyun Uhm}[]





\affiliation[1]{organization={Department of Electrical Engineering Korea University},
            state={Seoul},
            country={Korea}}

\author[1]{Seung-Won Jung}[]

\cormark[1]


\ead{swjung83@korea.ac.kr}



\affiliation[2]{organization={Department of Urology, The Catholic University of Korea},
            state={Seoul},
            country={Korea}}

\author[2]{Sung-Hoo Hong}[]

\author[3]{Sung-Jea Ko}[]

\affiliation[3]{organization={MedAI},
            state={Seoul},
            country={Korea}}
            
\cortext[1]{Corresponding author}



\begin{abstract}
Multi-phase computed tomography (CT) has been widely used for the preoperative diagnosis of kidney cancer due to its non-invasive nature and ability to characterize renal lesions. However, since enhancement patterns of renal lesions across CT phases are different even for the same lesion type, the visual assessment by radiologists suffers from inter-observer variability in clinical practice. Although deep learning-based approaches have been recently explored for differential diagnosis of kidney cancer, they do not explicitly model the relationships between CT phases in the network design, limiting the diagnostic performance. In this paper, we propose a novel lesion-aware cross-phase attention network (LACPANet) that can effectively capture temporal dependencies of renal lesions across CT phases to accurately classify the lesions into five major pathological subtypes from time-series multi-phase CT images. We introduce a 3D inter-phase lesion-aware attention mechanism to learn effective 3D lesion features that are used to estimate attention weights describing the inter-phase relations of the enhancement patterns. We also present a multi-scale attention scheme to capture and aggregate temporal patterns of lesion features at different spatial scales for further improvement. Extensive experiments on multi-phase CT scans of kidney cancer patients from the collected dataset demonstrate that our LACPANet outperforms state-of-the-art approaches in diagnostic accuracy.
\end{abstract}




\maketitle










\section{Introduction}\label{introduction}
Kidney cancer is one of the most common cancers in the world, with an estimated 76,080 newly diagnosed cases and 13,780 deaths in the United States in 2021~\cite{Cancer_Stats}. Approximately 90\% of all kidney cancers are renal cell carcinomas (RCCs)~\cite{epid_rcc}, and there are three main RCC types, \textit{i.e.}, clear cell RCC (ccRCC), papillary RCC (pRCC), and chromophobe RCC (chRCC), according to the 2016 World Health Organization (WHO) classification of renal tumors~\cite{WHO_2016}. As the prognosis of renal tumors depends on their histological subtype, 
the preoperative differential diagnosis of the tumors is essential for treatment planning~\cite{EAU_Guide}. 

\begin{figure}[!t]
\centering
\includegraphics[width=1.0\linewidth]{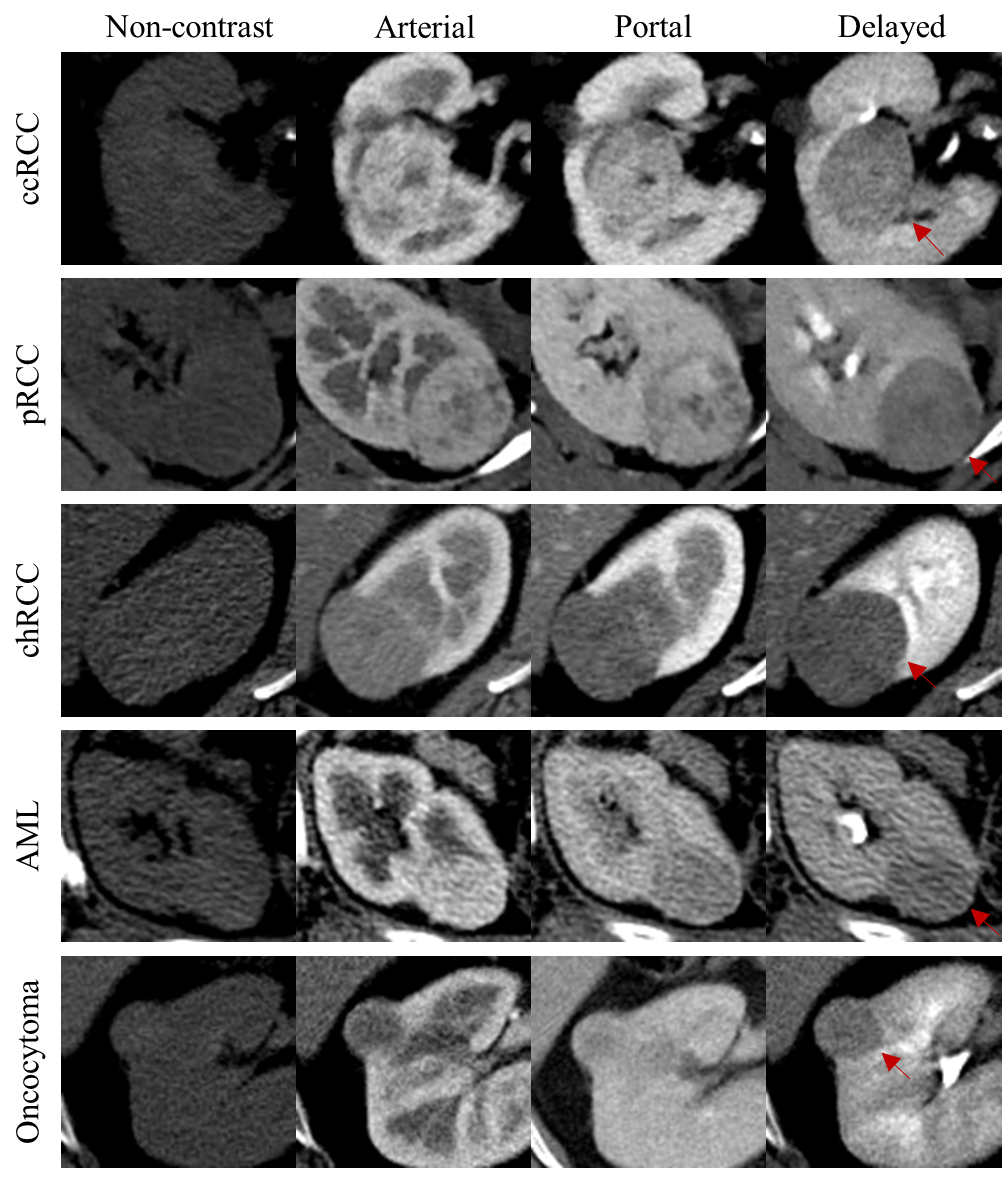}
\caption{
Sample of five subtypes of renal tumors in multi-phase CT scans. In each row, CT images of the corresponding subtype over multiple phases are shown. The tumor area is indicated by the red arrow in the delayed phase.
}
\label{fig:subtypes}
\end{figure}

Medical imaging techniques are widely used for the non-invasive diagnosis of renal tumors~\cite{imaging_rcc}, avoiding unnecessary biopsy. Multi-phase computed tomography (CT) has been considered the best diagnostic imaging modality for the detection and characterization of renal tumors due to its superiority over ultrasound and the limited availability of magnetic resonance imaging~\cite{NEJM_RCC, pic_review}. In multi-phase CT examination, a series of CT volumes are acquired before and after contrast injection at different time points. In particular, three contrast-enhanced phases, \textit{i.e.}, arterial, portal, and delayed phases, are typically acquired 20--30 s, 60--70 s, and 180 s after the injection, respectively. Radiologists then identify histologic subtypes of renal lesions by comparing the attenuation values and enhancement patterns of the lesions in the post-contrast images with their corresponding values in the non-contrast images~\cite{image_wisely} based on the studies \cite{Diff_RCC, ccrcc_mpct} that clear cell RCC usually leads to distinctive contrast enhancement in the arterial and delayed phases, and papillary and chromophobe RCCs are often substantially enhanced in the portal phase. Besides the degree of contrast enhancement, other lesion features, such as homogeneity of enhancement and calcification, are also used for the differentiation~\cite{MDCT_feat}. However, since renal tumors exhibit subtle differences in image features across cancer subtypes and the same type of renal lesions can exhibit variable enhancement patterns across CT phases, the visual assessment even by experienced radiologists suffers from considerable inter-observer variability~\cite{img_scr_kid}. 
Different subtypes of renal tumors acquired in the four phases are shown in Fig.~\ref{fig:subtypes}.
Moreover, benign renal lesions, such as fat-poor angiomyolipoma (AML) and oncocytoma, are frequently misdiagnosed as RCC, which may lead to unnecessary surgery~\cite{AML_diff}. Therefore, an automatic multi-phase CT analysis system is highly desired for the accurate and reliable diagnosis of renal lesions. 

With the recent advancement of deep convolutional neural networks (CNNs) in computer vision tasks, deep learning-based approaches have been applied to the differential diagnosis of renal lesions on multi-phase CT images~\cite{Tanaka_deep, review_urol_dl, inception_rcc, dl_3p_rcc, dl_4p_Oberai, npj_uhm}. Various CNN architectures commonly used in image classification, including AlexNet~\cite{AlexNet}, VGG~\cite{VGG}, GoogleNet~\cite{GoogleNet}, ResNet~\cite{ResNet}, and Inception-V3~\cite{InceptionV3}, have been explored for the classification of renal masses. To utilize multi-phase CT data, most existing methods adopt the early fusion strategy, which concatenates CT images from different phases along the channel dimension and feeds them to the network. However, since these methods do not explicitly model the temporal relationships between the enhancement patterns of lesions, which are crucial for classifying renal tumor subtypes, the previous approaches lead to sub-optimal performance, hindering their application in practice.

Recently, the self-attention mechanism in a transformer architecture, originally proposed for natural language processing (NLP)~\cite{Attn_all_u_need}, has been applied in various computer vision tasks, such as image recognition~\cite{ViT}, object detection~\cite{DETR}, and medical diagnosis~\cite{LAT_DR}. 
As a main building block of a transformer, it captures the contextual information from sequential data by computing attention scores between every pair of elements in the input sequence. For example, the input sequence can be a sequence of words in NLP or a stack of image patches in computer vision tasks~\cite{ViT}.
In the field of medical diagnosis, Sun \emph{et al.}~\cite{LAT_DR} used the self-attention mechanism to predict the severity of diabetic retinopathy from a given fundus image, where attention weights for lesions are obtained at each pixel. 
Despite the remarkable success of self-attention on sequential data, leveraging the merits of the attention mechanism 
for the classification of renal tumors in time-series multi-phase CT has not been actively explored in the literature.

In this paper, we propose a novel lesion-aware cross-phase attention network (LACPANet) that analyzes multi-phase CT images and classifies pathological subtypes of renal lesions into five major classes, including benign and malignant tumors, for accurate kidney cancer diagnosis. To explicitly learn temporal dependencies in enhancement patterns of lesions from different CT phases, LACPANet is embodied with a 3D inter-phase lesion-aware attention module that captures pairwise relations between CT phases based on 3D lesion features, which are generated by 3D CNNs with the guidance of the lesion segmentation network. This attention module encourages LACPANet to focus on more informative CT phases that are helpful for the differentiation of lesions in a given multi-phase CT. Further, we present a multi-scale version of the attention module to capture the temporal dependencies at multiple feature scales and aggregate the information from multiple scales for the final prediction, improving diagnostic performance over a single-scale approach. We extensively evaluate our LACPANet on the collected dataset that contains multi-phase CT scans of kidney cancer patients. Experimental results demonstrate that our method can achieve superior performance compared to the existing methods in fine-grained pathological subtype classification of renal lesions.

Our main contributions are summarized as follows:
\begin{itemize}
    \item To the best of our knowledge, this is the first work to leverage the attention mechanism of a transformer to analyze a time series of multi-phase CT images for the differential diagnosis of kidney cancer.
    \item We propose a 3D inter-phase lesion-aware attention mechanism to learn temporal dependencies between enhancement patterns of lesions from different CT phases based on 3D lesion features. 
    \item We present a multi-scale attention scheme to capture the dependencies at multiple feature scales and aggregate them for further improvement in diagnosis performance.
    \item Extensive experiments on the collected dataset demonstrate the effectiveness of our LACPANet for kidney cancer diagnosis in multi-phase CT.
\end{itemize}

The rest of this paper is organized as follows. In Section~\ref{sec:related_work}, related works on deep learning based-kidney cancer diagnosis and transformer are briefly reviewed. In Section~\ref{sec:method}, the proposed LACPANet is presented. Experimental validation against recent methods and analysis are described in Section~\ref{sec:experiments}. Conclusions are given in Section~\ref{sec:conclusion}.

\section{Related Work}\label{sec:related_work}

\subsection{Deep learning for kidney cancer diagnosis}
In recent years, deep learning has shown great potential in automated diagnosis of kidney cancer on abdominal CT scans. Coy \emph{et al.}~\cite{inception_rcc} utilize Inception-V3 architecture to discriminate clear cell RCC from oncocytoma by taking a single-phase CT with manually cropped tumor regions as input, obtaining the best classification result from the delayed phase. Zhou \emph{et al.}~\cite{radiomics} also use Inception-V3 to differentiate benign from malignant renal tumors based on single-phase CT images, where different CT attenuation ranges are used to analyze various tissue regions. 
Zabihollahy \emph{et al.}~\cite{decision_fusion} propose both semi-automated and fully-automated methods for differentiating RCC from benign renal masses on portal phase CT images by aggregating slice-level predictions, where the fully-automated approach involves U-Net~\cite{cicek2016unet3d} based segmentation network to segment renal masses automatically. They explore both 2D and 3D CNNs for the classification network.
Liu~\emph{et al.}~\cite{Dual_Input} present a dual-path classification network to utilize global and local features for the classification of five renal tumor subtypes, where global features are obtained from regions of renal mass with the surrounding kidney, while local features are extracted from only renal mass regions. 
Zhu \emph{et al.}~\cite{3DResNet_RCC} develop a 3D-based framework that crops ROI volumes using the result of the segmentation network and uses a pre-trained 3D-ResNet~\cite{3DResNet} for the tumor classification on single-phase CT. 
Kong~\emph{et al.}~\cite{BKC_Net} propose a 3D focus-perceive learning and domain knowledge embedding framework to mitigate the various degree of enhancement patterns of renal tumors and predict subtypes on single-phase CT. 

On the other hand, multi-phase CT-based approaches have also been explored. Han \emph{et al.}~\cite{dl_3p_rcc} combine three-phase CT images and then feed them into a CNN to distinguish three major subtypes of RCC. Oberai \emph{et al.}~\cite{dl_4p_Oberai} use four-phase CT images with cropped tumor regions as inputs to a CNN to classify malignant from benign lesions, where manually segmented tumor regions with the largest cross-sectional area in each CT phase are used as input to a model. 
Tanaka \emph{et al.}~\cite{Tanaka_deep} utilize Inception-v3~\cite{InceptionV3} model to determine if a small solid renal mass is benign (oncocytoma, AML) or malignant (ccRCC, pRCC, chRCC) on multi-phase contrast-enhanced CT, where each image is cropped around the lesion by an abdominal radiologist.
They compare single-phase-based approaches against multi-phase-based ones, where the best classification performance is obtained when the network is trained with arterial phase images alone. Uhm~\emph{et al.}~\cite{npj_uhm} propose an end-to-end framework that combines lesion segmentation and cancer subtype classification networks for fully automated diagnosis. However, these methods adopt the early-fusion strategy by concatenating the input sequence of CT phases. Consequently, they do not explicitly model the dependencies of enhancement patterns of lesions across CT phases that provide essential clues for characterizing renal lesions in multi-phase CT.

\begin{figure*}[!t]
\centering
\includegraphics[width=1.0\linewidth]{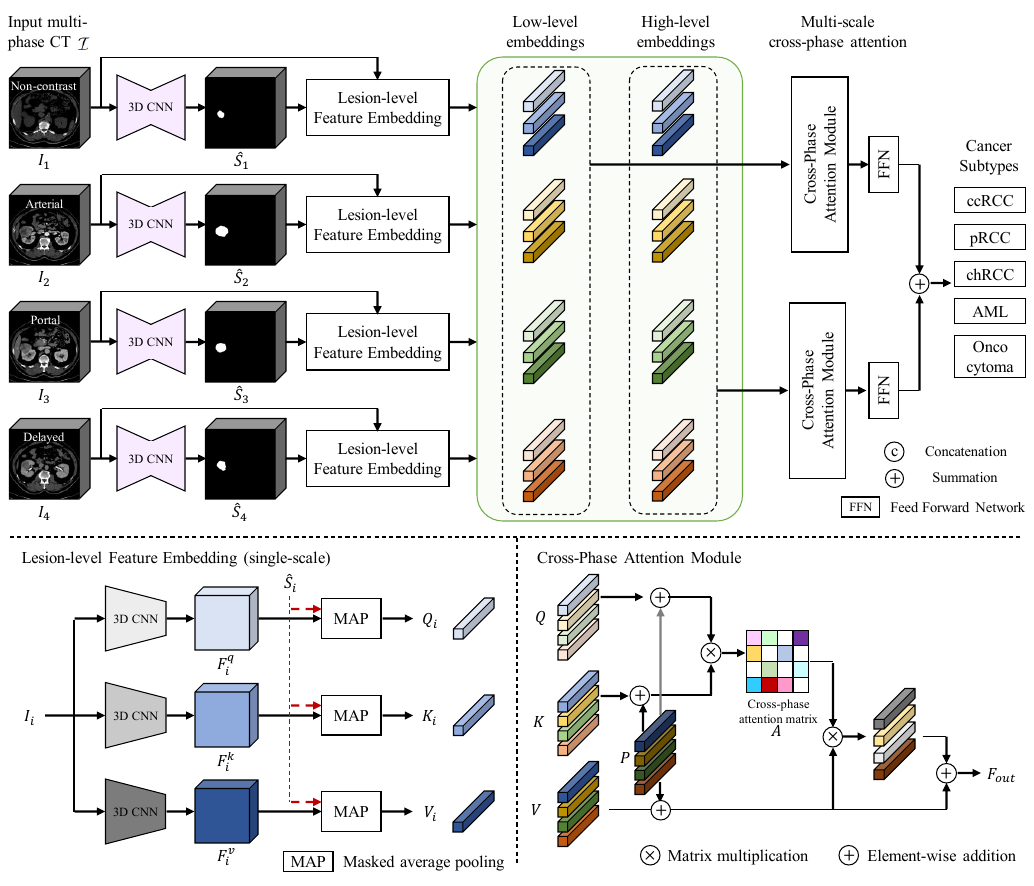}
\caption{Overall framework of our LACPANet. Given multi-phase CT $\mathcal{I}$, 3D CNN-based segmentation network predicts a lesion segmentation map $\hat{s}_{i}$ for each phase $I_i$. Then, lesion-level feature embedding (bottom left) is performed to produce lesion representations for each phase, \textit{i.e.}, query $Q$, key $K$, and value $V$. These embeddings are fed to the cross-phase attention module (bottom right) to capture inter-phase dependencies of lesion features with the help of phase embedding $P$. Here, only the single-scale process of lesion-level feature embedding was described for brevity. Finally, the output feature of cross-phase attention $F_{out}$ is fed to FFN to obtain the cancer subtype prediction.}
\label{fig:framework}
\end{figure*}

\subsection{Attention mechanism}

The attention mechanism is the core component of transformers, which was first introduced by Vaswani \emph{et al.}~\cite{Attn_all_u_need} and has been proven to be powerful in NLP tasks, replacing recurrent neural networks.
It can capture dependencies between elements of a sequence by aggregating global information from the whole sequence. 
Recently, with the success of transformers in NLP, there have been attempts to apply self-attention in various computer vision tasks such as image recognition~\cite{ViT}, object detection~\cite{DETR}, and medical diagnosis~\cite{LAT_DR, Drusen, Brain_age}. Dosovitskiy \emph{et al.}~\cite{ViT} introduce Vision Transformer (ViT), a pure transformer-based image classification model treating an input image as a sequence of 16$\times$16 patches, achieving state-of-the-art performance. Carion \emph{et al.}~\cite{DETR} propose Detection Transformer (DETR) for end-to-end object detection where models take an image feature sequence generated from a CNN to produce a final set of predictions. 

Recently, the self-attention mechanism of a transformer has also been explored in a wide variety of medical image analysis tasks, including disease diagnosis and other applications~\cite{Transformer_in_medical}.
Sun \emph{et al.}~\cite{LAT_DR} design a model for diabetic retinopathy grading from an input fundus image by introducing cross-attention between representative lesion filters and the pixels of the feature map. Wang \emph{et al.}~\cite{Drusen} present an attention-based architecture to perform drusen segmentation in retinal optical coherence tomography images for early diagnosis of age-related macular degeneration. He \emph{et al.}~\cite{Brain_age} propose a global-local transformer, which fuses local fine-grained details with the global-context information by the attention mechanism, for brain age estimation on magnetic resonance imaging.
For the diagnosis of COVID-19, the works described in~\cite{Trans_COVID_Peng, Trans_COVID_Tian} develop ensemble deep-learning models that combine transformer structure and CNNs to improve the classification performance.
Cao~\emph{et al.}~\cite{cancers15051538} introduce an attention-based model to predict microvascular invasion in hepatocellular carcinoma on preoperative contrast-enhanced CT.
However, leveraging the advantages of the attention mechanism of transformers in analyzing a time series of multi-phase CT for differential diagnosis of renal masses has not yet been explored. Therefore, in this work, we design a novel attention-based network to capture the temporal dependencies of contrast enhancement patterns of renal lesions across CT phases for accurate pathological subtype classification on multi-phase CT images.

\subsection{Multi-modal fusion for medical image analysis}
Multi-modal fusion has been widely explored in the analysis of medical images, especially for image segmentation tasks~\cite{SUN202134, ZHU2023376, tri-attention, mutual_learning, review_multimodal, M3Net, coco_attention}.
Some studies\cite{ZHU2023376, SUN202134} have leveraged multi-modal imaging data by adopting an early fusion approach that concatenates the modalities in the input space, but these methods do not exploit the relationships between modalities. To effectively exploit the multimodal information, cross-modal attention-based fusion strategies have been explored for the segmentation tasks~\cite{tri-attention, mutual_learning, M3Net, coco_attention}. 
Recently, for the classification task, Xu~\emph{et al.}~\cite{MCCNet} propose a framework that differentiates liver lesions by adaptively combining multi-phase information based on attention module. 
However, most previous cross-phase attention methods rely on 2D features for analyzing each 2D slice in 3D volumetric CT/MRI images, and thus cannot capture full 3D anatomy information of volumetric images.
Moreover, inter-modal attention is obtained via image-level features, which may not be optimal for lesion-level analysis.
To overcome these limitations, we propose a model that learns effective 3D features based on 3D lesion region analysis and lesion-aware attention mechanism to fully capture 3D lesion-level characteristics among CT phases and accurately classify lesion subtypes on multi-phase CT.

\section{Method}\label{sec:method}
In this section, we present a lesion-aware cross-phase attention network (LACPANet) for kidney cancer diagnosis. An overview of LACPANet is depicted in Fig. \ref{fig:framework}.
Given a sequence of multi-phase CT images, our goal is to learn a classifier that exploits dependencies between multiple phases in order to predict the pathological subtype of the tumor.
To explicitly model the dependencies of lesion features in the input CT sequence, we first propose a 3D inter-phase lesion-aware attention mechanism.
Then, we introduce the multi-scale attention scheme to capture the dependencies at multiple scales, which brings further improvement.
The rest of this section describes the details of each component.

\subsection{3D Inter-Phase Lesion-Aware Attention}\label{sec:IPLAT}
Let $\mathcal{I}=\{I_i\}_{i=1}^{N}$ be a set of images in a multi-phase CT scan, where $N$ is the number of CT phases and $I_i \in \mathbb{R}^{H \times W \times D}$ is the $i$-th image of the corresponding phase with resolution ($H$, $W$, $D$).
For example, $N=4$ if non-contrast, arterial, portal, and delayed phases are acquired during the scan.
We first extract a lesion segmentation map $\hat{S}_i\in\{0, 1\}^{H\times W \times D}$ for each image $I_i$ through a lesion segmentation network, where $1$ denotes the voxel belongs to a tumor area and $0$ otherwise.
Specifically, the lesion segmentation network is based on 3D CNNs detailed in Section~\ref{sec:Implementation}, and the weights of the network are shared across phases. 
Each $(I_i, \hat{S}_i)$ pair is then used for the lesion-level feature embedding to analyze the enhancement patterns of renal lesions in each phase $i$.
As illustrated in Fig. \ref{fig:framework} (bottom left), three separate networks produce three feature maps $F_i^q$, $F_i^k$, and $F_i^v \in \mathbb{R}^{H \times W \times D \times C}$, respectively, from the input image $I_i$, where $C$ denotes the number of channels. Each network consists of two $3\times3\times3$ convolutional layers with instance normalization~\cite{IN} and leaky ReLU activation.
To obtain lesion-level feature representations, the three feature maps are transformed into $Q_i$, $K_i$, and $V_i$ through masked average pooling (MAP)~\cite{SGOne} with the predicted segmentation mask $\hat{S}_i$:
\begin{equation}\label{eqa:QKV}
Q_i\!=\!\mathrm{MAP}(F_i^q, \hat{S}_i), K_i\!=\!\mathrm{MAP}(F_i^k, \hat{S}_i), V_i\!=\!\mathrm{MAP}(F_i^v, \hat{S}_i), 
\end{equation}
where $Q_i$, $K_i$, and $V_i \in \mathbb{R}^C$ and $\mathrm{MAP}(\cdot, \cdot)$ denotes the MAP operation, which is formulated as:
\begin{equation}\label{eqa:MAP}
{\rm{MAP}}\left( {F,\hat S} \right) = \left[{\left. {\frac{{\sum\limits_{{\bf{x}} \in \chi } {F\left( {{\bf{x}},c} \right)\hat S\left( \bf{x} \right)} }}{{\sum\limits_{\bf{x} \in \chi } {\hat S\left( \bf{x} \right)} }}} \right|,\,\,c = 1,2, \cdot  \cdot  \cdot ,C} \right],
\end{equation}
where ${\bf{x}}$ represents a 3D coordinate, and $\mathcal{X}$ is the set of all 3D spatial locations of $F$ and $\hat{S}$. We use the extracted embeddings $Q_i$, $K_i$, and $V_i$ as query, key, and value, respectively, to capture the inter-phase relationships for cancer subtype classification.

After the lesion-level feature embeddings are obtained, phase embeddings are added to queries, keys, and values, as can be seen in Fig. \ref{fig:framework}.
We first define the set of 1D learnable phase embeddings ${P}=\{P_i\}_{i=1}^N$, where $P_i \in \mathbb{R}^{C}$ is the phase embedding for the $i$-th phase.
Then, the phase embedding $P_i$ is added to the query $Q_i$, key $K_i$, and value $V_i$ to provide information indicating which phase these feature embeddings belong to.

Next, we model inter-phase dependencies by exploiting the self-attention mechanism of transformers~\cite{Attn_all_u_need}, as shown in Fig.~\ref{fig:framework} (bottom right).
We denote the query, key, and value matrices as $Q$, $K$, and $V \in \mathbb{R}^{N \times C}$, respectively. Specifically, the $i$-th rows of $Q$, $K$, and $V$ are given as ${Q_i} + {P_i}$, ${K_i} + {P_i}$, and ${V_i} + {P_i}$, respectively. 
The cross-phase attention module first computes the scaled dot products between $Q$ and $K$ and applies a softmax function to obtain an attention weight matrix $A \in \mathbb{R}^{N \times N}$, which indicates the interdependence of pairwise lesion features from different phases, which can be expressed as:
\begin{equation}\label{eqa:attention}
A(Q, K)=\mathrm{softmax}\left(\frac{QK^T}{\sqrt{C}}\right).
\end{equation}
The scaled dot product between $Q$ and $K$ computes the similarity of the features across different CT scans.
For example, $Q_iK_J^T$ measures the similarity between the query feature of the $i^{th}$ phase and the key feature of the $j^{th}$ phase.
Thus, the attention matrix $A$ represents the interdependencies of the lesion features of different CT phases.
Then, the value matrix $V$ is combined with its multiplication with the attention weight matrix $A$ to produce the output feature matrix $F_{out} \in \mathbb{R}^{N \times C}$: 
\begin{equation}\label{eqa:residual}
F_{out} = \lambda AV + V,
\end{equation}
where $\lambda$ is a weighting hyper-parameter, setting as $\lambda = 0.1$ empirically. 
Through this process, the lesion features for different CT phases in $V$ are weighted according to the attention matrix $A$, enhancing the features by reflecting interdependencies between CT phases.
We then reshape $F_{out}$ to obtain a 1D fused feature vector $f_{out} \in \mathbb{R}^{NC}$. The probability prediction $\hat{y}$ for cancer subtype classification is produced by a feed-forward network (FFN) taking the output feature vector $f_{out}$ as input, \textit{i.e.},  $\hat{y}=\mathrm{softmax}(\mathrm{FFN}(f_{out}))$, where $\hat{y}$ represents probability distribution over cancer subtype classes.
We note that $\hat{y}$ is the prediction result at single-scale.
The final output of LACPANet is obtained through a multi-scale attention scheme described in Section~\ref{sec:MSA}.

\begin{figure}[!t]
\centering
\includegraphics[width=1.0\linewidth]{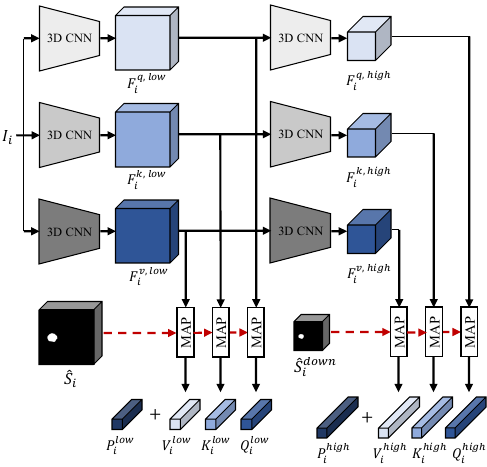}
\caption{Multi-scale attention scheme. We use both low-level and high-level features produced by 3D CNNs to compute inter-phase attention. The multi-scale lesion-level embeddings are obtained with the 3D features and the segmentation maps at the corresponding scales. These feature embeddings are then added by learnable phase embeddings at multiple scales. }
\label{fig:multi_scale}
\end{figure}

\subsection{Multi-Scale Attention }\label{sec:MSA}
For differential diagnosis of cancer subtypes, analyzing CT image features of tumors, such as textures and structures of lesions, at various scales rather than single-scale can be beneficial~\cite{multi_scale_frontier}. 
We therefore introduce a multi-scale attention scheme, which captures inter-phase dependencies of lesion features at different scales to promote our model to more accurately produce predictions, as shown in Fig.~\ref{fig:multi_scale}.

We start by extracting multi-scale deep features from three separate encoders.
Let $F_{i}^{q, low}$, $F_{i}^{k, low}$, and $F_{i}^{v, low}\in\mathbb{R}^{H \times W \times D \times C}$ be the low-level features produced by the first two layers of each encoder, and $F_{i}^{q, high}$, $F_{i}^{k, high}$, and $F_{i}^{v, high} \in \mathbb{R}^{\frac{H}{2} \times \frac{W}{2} \times \frac{D}{2} \times 2C}$ be the high-level features from the output of each encoder.
The components of the first two layers of the encoder are the same as described in Sec.~\ref{sec:IPLAT}, while the other two appended 3D convolutional layers (the first layer performs $3\times3\times3$ convolution with stride 2 for the downsampling) are used to extract high-level features. We share the weights of the encoders across all phases.

We then down-sample the predicted segmentation map $\hat{S}_i$ to be of size $\frac{H}{2} \times \frac{W}{2} \times \frac{D}{2}$ to match the spatial resolution of features and segmentation maps.
We denote the down-sampled segmentation map as $\hat{S}_{i}^{down}$. 
After that, we convert feature maps at each scale into queries, keys, and values through MAP with the segmentation map at the corresponding scale.
The low-level feature embeddings, $Q_i^{low}$, $K_i^{low}$, and $V_i^{low}\in\mathbb{R}^C$ are obtained using $\hat{S}_i$, and the high-level feature embeddings, $Q_i^{high}$, $K_i^{high}$, and $V_i^{high}\in\mathbb{R}^{2C}$, are computed using the down-sampled segmentation map $\hat{S}_{i}^{down}$.
We define the two sets of 1D learnable phase embeddings $\{P_i^{low}\in\mathbb{R}^{C}\}_{i=1}^N$ and $\{P_i^{high}\in\mathbb{R}^{2C}\}_{i=1}^N$ for low-level and high-level embeddings, respectively. The phase embeddings at each scale are then added to queries, keys, and values to retain positional information of the phases. 

Next, at each scale, the inter-phase dependencies of lesion features are captured through the cross-phase attention module, producing low-level and high-level attention matrices, $A^{low}$ and $A^{high}$. The outputs of the cross-phase attention at low-level and high-level are thus the feature matrices $F_{out}^{low}\in\mathbb{R}^{N\times C}$ and $F_{out}^{high}\in\mathbb{R}^{N\times 2C}$, which are reshaped to form the vectors $f_{out}^{low}\in\mathbb{R}^{NC}$ and $f_{out}^{high}\in\mathbb{R}^{2NC}$.
These feature vectors are used to make low-level and high-level cancer subtype predictions, \textit{i.e.,} $\hat{y}^{low}$ and $\hat{y}^{high}$, which can be expressed as follows:

\begin{equation}\label{eqa:low_high}
\begin{aligned}
\hat{y}^{low} = \mathrm{softmax}(\mathrm{FFN}(f_{out}^{low})),  \\
\hat{y}^{high} = \mathrm{softmax}(\mathrm{FFN}(f_{out}^{high})). \\
\end{aligned}
\end{equation}
The final cancer subtype prediction result $\hat{y}^{final}$ is obtained as a weighted average between both predictions:

\begin{equation}\label{eqa:y_final}
\hat{y}^{final} = \alpha\hat{y}^{low} + (1-\alpha)\hat{y}^{high},
\end{equation}
where $\alpha$ is a hyper-parameter for balancing low and high-level predictions.

Finally, the training of our multi-scale model is supervised with the overall classification loss $\mathcal{L}$ on all scales, which is formulated as follows:
\begin{equation}\label{eqa:loss_multi}
\mathcal{L} = \mathcal{L}_{CE}(\hat{y}^{low}, y) + \beta \mathcal{L}_{CE}(\hat{y}^{high}, y),
\end{equation}
where $\mathcal{L}_{CE}$ is the cross-entropy loss between predictions and ground-truth subtype labels. $\beta$ is a hyper-parameter to balance the two loss terms.

\section{Results}\label{sec:experiments}

In this section, we first describe our dataset in Section~\ref{sec:dataset} and provide implementation details of our proposed model in Section~\ref{sec:Implementation}. 
Then, we introduce baseline methods and evaluation metrics in Section~\ref{sec:baselines}.
Next, we present quantitative comparisons against the existing methods and ablation studies on key components of our model in Section~\ref{sec:quantitative}.
The ablation studies for hyper-parameter selection are presented in Section~\ref{sec:ablation}.
We visualize the attention responses of our model in~Section~\ref{sec:visualize_attention}. 


\begin{table}[!t]
\caption{Patient demographics and clinical characteristics of the study.}
\centering
\begin{tabular}{|l|c|c|}
\hline
\textbf{Total patients ($n$)}& 183 \\
\hline 
Age (years) & 54.79 $\pm$ 11.98\\
Gender (M/F) & 104 / 79 \\ 
Tumor size (cm) & 3.34 $\pm$ 2.15 \\ 
\hline 
\textbf{Subtype} &  \\
ccRCC & 62\\ 
pRCC & 36\\ 
chRCC & 34 \\ 
AML & 31 \\ 
Oncocytoma & 20 \\ 

\hline
\end{tabular}
\label{tab:demographics}
\end{table}

\subsection{Dataset}\label{sec:dataset}
Due to the lack of publicly available datasets for multi-phase CT-based kidney cancer subtype classification, we utilize the dataset collected from Seoul St. Mary's Hospital as described in our previous studies~\cite{npj_uhm, 2022_uhm}. 
This study was approved by the Seoul St. Mary's Hospital Institutional Review Board.
The dataset consists of preoperative 3D multi-phase dynamic contrast-enhanced CT scans from kidney cancer patients who underwent nephrectomy for renal tumors at Seoul St. Mary's Hospital and underwent abdominal CT at Seoul St. Mary's Hospital or other hospitals three months before surgery.
The dataset contains five major subtypes of renal tumor (ccRCC, pRCC, chRCC, AML, and oncocytoma), where all renal tumor subtypes are pathologically confirmed by surgical resection. 
In the data collection process, ccRCC samples with large tumor sizes ($>7$cm) were excluded as they can be easily diagnosed by radiologists in practice.
CT images were captured by models from various CT imaging manufacturers, including GE, Hitachi, Philips, SIEMENS, and TOSHIBA.
We refer the reader to the supplementary material in \cite{npj_uhm} for the detailed distributions of manufacturers and model names of CT scanners.
The spatial resolution of CT images is 512×512.
The slice thickness varies from 1 to 7mm, and pixel spacing ranges from 0.53 to 0.94mm.
We use 183 complete four-phase CT scans (732 3D volumes) to train and test our model, where the four-phase CT is composed of non-contrast, arterial, portal, and delayed phases.
Table~\ref{tab:demographics} shows the demographic and clinical characteristics of the study population. 
The representativeness of the dataset was justified by showing that the model
trained on this dataset generalized well to other publicly available datasets for the three-class subtype classification~\cite{npj_uhm} and the usefulness of the dataset was also verified in
another task of CT image synthesis~\cite{2022_uhm}.

We generate train/val/test splits with a ratio of 65\%:15\%: 20\%.  
In the test split, the number of cases for each subtype is 9, 7, 7, 7, 7 for ccRCC, pRCC, chRCC, AML, and oncocytoma, respectively.
The ground-truth tumor segmentation labels for individual CT phases are first manually annotated by trained annotators and then checked and reﬁned an experienced radiologist.
We preprocess CT data by resampling the volumes to $1.5\times1.5\times3\mathrm{mm}^3$ voxel size, truncating intensity values to [-40, 350] HU followed by intensity normalization, and cropping $192\times160\times96$ regions, including kidneys and tumors, for computation efficiency.

\subsection{Implementation Details}\label{sec:Implementation}
We implement our approach using PyTorch~\cite{Pytorch} and train our model on an NVIDIA TITAN XP GPU with 12GB memory. We adopt 3D U-Net architecture~\cite{cicek2016unet3d} for the lesion segmentation network, where $3\times3\times3$ convolutional layers are followed by instance normalization~\cite{IN} and LeakyReLU~\cite{LeakyRelu} activation.
We use 3D convolution layers with a stride of 2 to downsample features and transposed convolutions with stride 2 for upsampling operation. 
The segmentation network is trained using a Dice loss~\cite{Dice} with stochastic gradient descent.
The FFN of three fully connected layers is employed to produce probability distribution over subtype classes for the classifier. 
We train our LACPANet with $\mathcal{L}$ for 200 epochs with a mini-batch size of 1 and an initial learning rate of $10^{-4}$, which is reduced by a factor of $10^{-1}$ every 50 epochs. Random flipping along the x-, y-, and z-axis with a probability of 0.5 for each axis is used for data augumentation in the training phase.
We utilize ADAM optimizer~\cite{Adam} with $\beta_1$=0.9, $\beta_2$=0.999, and $\epsilon$=$10^{-8}$. We set $C=8$, $\lambda=0.1$ in (\ref{eqa:residual}), $\alpha=0.7$ in (\ref{eqa:y_final}), and $\beta=0.1$ in (\ref{eqa:loss_multi}).
We initialize phase embeddings $P$ with random initialization, which samples from a zero-mean Gaussian distribution with standard deviation of $\sqrt{C}$.
\begin{figure}[t]
\centering
\includegraphics[width=1.0\linewidth]{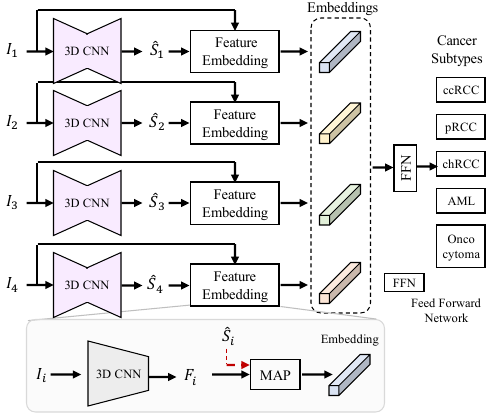}
\caption{Illustration of the 3D multi-phase baseline network. }
\label{fig:multi_scale}
\end{figure}

\subsection{Baselines}\label{sec:baselines}
We compare our proposed LACPANet with several existing deep learning-based renal tumor subtype classification methods.
We include both single-phase-based methods and multi-phase-based methods for comprehensive comparisons.
There are semi-automated methods that require manually annotated lesion regions of interest (ROIs)~\cite{dl_3p_rcc, dl_4p_Oberai, inception_rcc, Tanaka_deep, decision_fusion} and fully-automated methods that automatically segment renal mass regions and classify the subtypes of the mass in an end-to-end manner~\cite{decision_fusion, npj_uhm, 3DResNet_RCC, Dual_Input, BKC_Net}.
For implementing semi-automated baselines, we crop the boundary boxes of the lesions as ROIs using the pixel-level ground-truth segmentation maps and then feed cropped images to the models.

Most existing methods performed binary or 3-class classification rather than the five-class classification. 
Thus we modified and trained them to perform five-class classification on our dataset.
Since the five-class classification is much more challenging than binary classification, their performance may be lower than that of the original paper.
We note that differentiating five renal tumor subtypes solely on CT is challenging even for experienced radiologists, as reported in~\cite{npj_uhm}.
In addition, because our dataset mainly includes small-sized (3.34cm on average in Table~\ref{tab:demographics}) tumors that are much more difficult to differentiate, the overall performance of the models may be low compared to the original paper.
It is worth mentioning that we utilize the same experimental setting for all the methods to ensure fairness.

Although there exist many prior works on deep learning-based renal tumor classification, these methods have some limitations in that they use only single-phase CT as input or are based on 2D CNNs for classification. Therefore, we design a baseline for 3D multi-phase CT, which uses multi-phase CT as input and utilizes 3D CNNs to process volumetric CT data. Specifically, this baseline is constructed by discarding both the 3D inter-phase lesion-aware attention module and multi-scale attention from our LACPANet. We evaluate our method in two settings, semi-automated and fully-automated settings. Our LACPANet described in Section~\ref{sec:method} is used for the fully-automated setting.
For the semi-automated setting of our LACPANet, the pixel-level ground-truth segmentation map $S_i$ is used directly in the MAP module.


\subsection{Evaluation Metrics}\label{sec:metrics}
For performance comparison of classification models, we adopt the area under the curve (AUC) of the receiver operating characteristics (ROC) curve as the evaluation metric.
Since the ROC curve is based on the true positive rate (TPR) and the false positive rate (FPR), we obtain a ROC curve for each class by computing TPR and FPR using a one-vs-rest strategy, resulting in five ROC curves for different subtypes. We use the averaged AUC score weighted by support over classes for the comparison. The AUC value of 0.5 corresponds to random guessing, and AUC values close to 1 indicate better performance. We also use other popular metrics, \textit{i.e.}, precision, recall, and Fl score, to evaluate the performance of methods, where F1 score is the harmonic mean of precision and recall: $\frac{2}{\mathrm{recall}^{-1} + \mathrm{precision}^{-1}}$.
The precision, recall, and F1 scores are calculated for each subtype label, and then their average values weighted by support are computed.  

\begin{table*}
\caption{Quantitative comparison of cancer subtype classification performances under the semi-automated setting.}
\begin{center}
\begin{tabular}{ l c c c c c c c }
\hline 
Method & Input phase & 2D/3D & AUC $\uparrow$ & Precision $\uparrow$ & Recall $\uparrow$ & F1 score $\uparrow$ \\
\hline
Han \emph{et al.}~\cite{dl_3p_rcc} & Multi-phase & 2D & 0.8708 & 0.6216 & 0.6216 & 0.5830  \\
Oberai \emph{et al.}~\cite{dl_4p_Oberai} & Multi-phase & 2D & 0.7993 & 0.6185 & 0.5405 & 0.5475  \\
\hline
\multirow{2}{10em}{Coy \emph{et al.}~\cite{inception_rcc}} & Single-phase & \multirow{2}{*}{2D} & 0.8205 & 0.6184 & 0.5676 & 0.5606  \\
 & Multi-phase &  & 0.8201 & 0.5486 & 0.5946 & 0.5653  \\
\hline
\multirow{2}{10em}{Tanaka \emph{et al.}~\cite{Tanaka_deep}} & Single-phase & \multirow{2}{*}{2D} & 0.9086 & 0.6931 & 0.7027 & 0.6864  \\
 & Multi-phase & & 0.8963 & 0.5882 & 0.6486 & 0.5767  \\
\hline
\multirow{2}{10em}{Zabihollahy \emph{et al.}~\cite{decision_fusion}} & \multirow{2}{*}{Single-phase} & 2D & 0.8866 & 0.6388 & 0.6216 & 0.5966  \\
 & & 3D & 0.8973 & 0.7236 & 0.6216 & 0.6065  \\
\hline
3D multi-phase baseline & Multi-phase & 3D & 0.9186  & 0.6349 & 0.7027 & 0.6635 \\
+ our lesion-aware attention w/o $P$ & Multi-phase & 3D & 0.9249 & 0.6399 & 0.7027 & 0.6650 \\
Ours w/o multi-scale attention & Multi-phase & 3D & 0.9395  & 0.6805 & 0.7568 & 0.7146 \\
Ours (LACPANet) & Multi-phase & 3D & \textbf{0.9426}  & \textbf{0.8511} & \textbf{0.8108} & \textbf{0.7979} \\
\hline
\end{tabular}
\end{center}
\label{tab:comparison_gt}
\end{table*}

\begin{table*}
\caption{Quantitative comparison of cancer subtype classification performances under the fully-automated setting.}
\begin{center}
\begin{tabular}{ l c c c c c c c }
\hline 
Method & Input phase & 2D/3D & AUC $\uparrow$  & Precision $\uparrow$ & Recall $\uparrow$ & F1 score $\uparrow$ \\
\hline
Zabihollahy \emph{et al.}~\cite{decision_fusion} & Single-phase & 2D & 0.8535  & 0.5637 & 0.5946 & 0.5622 \\
Uhm \emph{et al.}~\cite{npj_uhm} & Multi-phase & 2D & 0.8251  & 0.6130 & 0.6486 & 0.6116 \\
Zhu \emph{et al.}~\cite{3DResNet_RCC} & Single-phase & 3D & 0.8294  & 0.6213 & 0.5135 & 0.4508 \\
Liu \emph{et al.}~\cite{Dual_Input} & Single-phase & 2D & 0.8484 & 0.6179 & 0.6486 & 0.6082 \\
Kong \emph{et al.}~\cite{BKC_Net} & Single-phase & 3D & 0.8237  & 0.5304 & 0.5405 & 0.4968 \\
\hline
3D multi-phase baseline & Multi-phase & 3D & 0.8440 & 0.5429 & 0.5946 & 0.5584 \\
+ our lesion-aware attention w/o $P$ & Multi-phase & 3D & 0.8645 & 0.6110 & 0.6216 & 0.5959 \\
Ours w/o multi-scale attention & Multi-phase & 3D & 0.8807 & 0.7053 & 0.6757 & 0.6621 \\
Ours (LACPANet) & Multi-phase & 3D & \textbf{0.9022} & \textbf{0.7707} & \textbf{0.7568} & \textbf{0.7536} \\
\hline
\end{tabular}
\end{center}
\label{tab:comparison_seg}
\end{table*}

\begin{table*}
\caption{Comparison with existing multi-phase fusion approaches under the semi-automated and fully-automated setting.}
\begin{center}
\begin{tabular}{ l c c c c  c c c c}
\hline 
\multirow{2}{*}{Method} &\multicolumn{4}{c}{Semi-automated} &\multicolumn{4}{c}{Fully-automated}\\
\cmidrule(lr){2-5} \cmidrule(lr){6-9}
 & AUC $\uparrow$ & Precision $\uparrow$ & Recall $\uparrow$ & F1 score $\uparrow$ & AUC $\uparrow$ & Precision $\uparrow$ & Recall $\uparrow$ & F1 score $\uparrow$\\
\hline
3D Baseline & 0.9186  & 0.6349 & 0.7027 & 0.6635 & 0.8440 & 0.5429 & 0.5946 & 0.5584 \\
Early fusion & 0.8974 & 0.6257 & 0.6757 & 0.6486 & 0.8277 & 0.5690 & 0.5405 & 0.5790  \\
MAM~\cite{tri-attention} & 0.9210 & 0.6632 & 0.7297 & 0.6898 & 0.8320 & 0.5284 & 0.5676 & 0.5312  \\
PA block~\cite{PAResSeg} & 0.8853 & 0.6385 & 0.6118 & 0.6757 & 0.8664 & 0.6574 & 0.6216 & 0.6087  \\
MA module~\cite{mutual_learning} & 0.9026 & 0.6433 & 0.7027 & 0.6670 & 0.8674 & 0.5333 & 0.5946 & 0.5565  \\
CPNL module~\cite{M3Net} & 0.8659 & 0.5616 & 0.6216 & 0.5862 & 0.8542 & 0.5886 & 0.6486 & 0.6136  \\
MSA module\cite{MCCNet} & 0.9177 & 0.6304 & 0.6757 & 0.6314 & 0.8574 & 0.5935 & \textbf{0.6757} & 0.6490  \\
\hline
Ours (single-scale) & \textbf{0.9395}  & \textbf{0.6805} & \textbf{0.7568} & \textbf{0.7146} & \textbf{0.8807} & \textbf{0.7053} & \textbf{0.6757} & \textbf{0.6621}  \\
\hline
\end{tabular}
\end{center}
\label{tab:comparison_att}
\end{table*}

\subsection{Quantitative Comparison}\label{sec:quantitative}
We first compare our LACPANet with the existing semi-automated methods~\cite{dl_3p_rcc, dl_4p_Oberai, inception_rcc, Tanaka_deep, decision_fusion} under the setting where the manual annotations are given.
Then, we compare our model with state-of-the-art fully-automated methods~\cite{decision_fusion, npj_uhm, 3DResNet_RCC, Dual_Input, BKC_Net} that predict cancer subtypes from input CT data in an end-to-end manner.

Table~\ref{tab:comparison_gt} shows the comparison results for the semi-automated baseline methods. Note that we evaluate our LACPANet under the semi-automated setting, as described in Section~\ref{sec:baselines}, to make a fair comparison.
As shown in the table, the proposed LACPANet outperforms the other models by a large margin in all evaluation metrics, demonstrating its effectiveness for cancer subtype classification from multi-phase CT data. We observe that our 3D multi-phase baseline already achieves better performance than the other previous methods with respect to AUC, Recall, and F1 score, which illustrates that using multi-phase CT and 3D CNNs is vital to extract discriminative features for subtyping renal tumors.
It can also be seen from the results of the method~\cite{decision_fusion} that the 3D variant shows better performance over the 2D variant, indicating the ability of 3D CNNs for analyzing 3D lesion volumes.    
In addition, in the results of the previous methods~\cite{inception_rcc, Tanaka_deep}, the multi-phase variants do not improve performance over the single-phase variants mostly due to their simple early fusion strategy for multi-phase input.

Table~\ref{tab:comparison_gt} also provides results that validate the effectiveness of different modules included in the proposed LACPANet for classifying subtypes. The 3D multi-phase baseline network equipped with our 3D inter-phase lesion-aware attention mechanism (denoted as ``Ours w/o multi-scale attention'') achieves performance improvement around 2.09\%, 4.56\%, 5.41\%, and 5.11\% in terms of AUC, precision, recall, and F1 score, respectively. 
The major reason for the performance improvement is that while the baseline extracts the features of each phase independently, the features in our model have been enhanced by leveraging inter-phase relationships to help the classification of lesions.
This demonstrates that our inter-phase lesion-aware attention is beneficial for learning multi-phase lesion features by capturing inter-dependencies across phases.
If the phase embedding $P$ in our lesion-aware cross-phase attention module is not used (denoted as ``+ our lesion-aware attention w/o $P$''), the degree of performance improvement becomes smaller, which shows the effectiveness of the proposed phase embedding $P$ in our network.
Moreover, further integrating our multi-scale attention scheme into the network increases the classification performance by 0.31\%, 17.1\%, 5.4\%, and 8.33\% in terms of AUC, precision, recall, and F1 score, respectively, which shows that capturing inter-phase dependencies at multiple scales significantly enriches the visual features of renal masses for differentiating subtypes. 
This appears to be because the effective feature scale for subtype classification may vary depending on the renal tumor.

To validate our proposed attention scheme over other multi-modal fusion strategies, We compare our method with existing multi-phase CT fusion approaches, including early fusion, modality-aware module (MAM)~\cite{tri-attention}, phase attention (PA) block~\cite{PAResSeg}, modality attention (MA) module~\cite{mutual_learning}, cross-phase non-local attention (CPNL) module~\cite{M3Net}, and multi-phase self-attention (MSA) module~\cite{MCCNet}, by adapting them to the baseline architecture.
The results are shown in Table~\ref{tab:comparison_att}. We see that the proposed method shows much improvement compared to the previous approaches. 
LACPANet is able to get better results by using our lesion-aware cross-phase module with learnable phase embeddings.

In Table~\ref{tab:comparison_seg}, we compare our model with state-of-the-art renal subtype classification methods under the fully automated setting that does not require manual annotation of lesion ROIs. As shown in the table, our LACPANet significantly outperforms all baseline methods for all evaluation metrics, benefiting from exploring the multi-phase CT information and analyzing 3D lesion volumes.
For example, the AUC of LACPANet is 0.9022, while that of the second-best method \cite{decision_fusion} is 0.8535. This is due to the fact that the previous methods cannot leverage information across CT phases or volumetric 3D representation. 
The comparison with existing multi-phase fusion approaches under the semi-automated setting is shown in Table~\ref{tab:comparison_att}. we find that the previous approaches yield less performance gain against our lesion-aware attention.

Also, we conduct ablation studies to validate the effectiveness of each component in the proposed model. 
The 3D multi-phase baseline model yields 0.8440, 0.5429, 0.5946, and 0.5584 in AUC, precision, recall, and f1-score.
We can observe that ``Ours w/o multi-scale attention'' boosts performance over the 3D multi-phase baseline around 3.7\%, 16.2\%, 8.1\%, and 10.4\% in terms of AUC, precision, recall, and F1 score, respectively. This result indicates the benefit of the proposed inter-phase attention module to fuse information from multiple CT phases for the subtype classification. 
Note that the model equipped with our attention module without the proposed phase embedding $P$ (``+ our lesion-aware attention w/o $P$'') achieves a smaller performance gain, which validates the necessity of our learnable phase embedding $P$.
In addition, when we apply the proposed multi-scale attention scheme to the model, our model continuously achieves improvements of 1.15\%, 6.54\%, 8.11\%, and 9.15\% for AUC, precision, recall, and F1 score, respectively.
This result suggests the effectiveness of the multi-scale version of our lesion-aware cross-attention in capturing inter-phase dependencies at multiple scales for differential diagnosis of renal tumors under the fully-automated setting.
In the ablation experiments for our model, the cause of the performance improvement is considered to be the same as described in the previous semi-automated setting experiment results.

Note that a semi-automatic approach is ideal in that ground-truth segmentation is used for classification; thus, its result serves as a performance upper bound.
Enhancing the performance of the segmentation network $S$ may benefit the classification task, but in this paper, the focus is not on proposing a new segmentation method. 
Instead, we use a top-performing 3D U-Net model, which shows its superiority throughout successive kidney and kidney tumor segmentation challenges~\cite{kits19, kits21, kits23_khuhm}.
We note that on the test cases, the average Dice scores of the segmentation network for kidney and tumor are 0.969$\pm$0.014 and 0.856$\pm$0.131, respectively.  
Overall, quantitative comparison results demonstrate that our method achieves superior performance over all baselines in both semi- and fully-automated settings.

\begin{table*}
\caption{Ablation study of the hyper-parameter $\lambda$. $\lambda=0.1$ is adopted as the default setting in other experiments.}
\begin{center}
\begin{tabular}{ l c c c c c c c c c }
\hline 
\multirow{2.5}{*}{$\lambda$} & \multicolumn{4}{c}{Semi-automated} & \multicolumn{4}{c}{Fully-automated} \\
\cmidrule(lr){2-5} \cmidrule(lr){6-9} 
 & AUC $\uparrow$  & Precision $\uparrow$ & Recall $\uparrow$ & F1 score $\uparrow$ & AUC $\uparrow$  & Precision $\uparrow$ & Recall $\uparrow$ & F1 score $\uparrow$\\
\hline
0.01 & 0.8885  & 0.6234 & 0.6757 & 0.6440 & 0.7649  & 0.4054 & 0.4595 & 0.4203 \\
0.1 & \textbf{0.9395}  & \textbf{0.6805} & \textbf{0.7568} & \textbf{0.7146} & \textbf{0.8713}  & \textbf{0.5766} & \textbf{0.6486} & \textbf{0.6045} \\
1.0 & 0.8969  & 0.6271 & 0.6757 & 0.6362 & 0.7732  & 0.4141 & 0.4865 & 0.4303 \\
\hline
\end{tabular}
\end{center}
\label{tab:abl_lambda}
\end{table*}

\begin{table*}[t]
\caption{Ablation study of the hyper-parameter $\alpha$. $\alpha=0.7$ is adopted as the default setting in other experiments.}
\begin{center}
\begin{tabular}{ l c c c c  c c c c }
\hline 
\multirow{2.5}{*}{$\alpha$} & \multicolumn{4}{c}{Semi-automated} & \multicolumn{4}{c}{Fully-automated} \\
\cmidrule(lr){2-5} \cmidrule(lr){6-9} 
 & AUC $\uparrow$  & Precision $\uparrow$ & Recall $\uparrow$ & F1 score $\uparrow$ & AUC $\uparrow$  & Precision $\uparrow$ & Recall $\uparrow$ & F1 score $\uparrow$\\
\hline
0.1 & 0.9068  & 0.6776 & 0.6216 & 0.6269 & 0.8878  & 0.7230 & 0.7297 & 0.7195 \\
0.3 & 0.9222  & 0.7139 & 0.6486 & 0.6515 & 0.8970  & 0.7230 & 0.7297 & 0.7195 \\
0.5 & 0.9359  & 0.7836 & 0.7568 & 0.7481 & 0.8958  & 0.7230 & 0.7297 & 0.7195 \\
0.7 & \textbf{0.9426} & \textbf{0.8511} & \textbf{0.8108} & \textbf{0.7979} & \textbf{0.9022} & \textbf{0.7707} & \textbf{0.7568} & \textbf{0.7536} \\
0.9 & 0.9383  & 0.6805 & 0.7568 & 0.7146 & 0.9006  & 0.5901 & 0.6486 & 0.6136 \\
\hline
\end{tabular}
\end{center}
\label{tab:abl_alpha}
\end{table*}

\begin{table*}[t]
\caption{Ablation study of the hyper-parameter $\beta$. $\beta=0.1$ is adopted as the default setting in other experiments.}
\begin{center}
\begin{tabular}{ l c c c c c c c c c }
\hline 
\multirow{2.5}{*}{$\beta$} & \multicolumn{4}{c}{Semi-automated} & \multicolumn{4}{c}{Fully-automated} \\
\cmidrule(lr){2-5} \cmidrule(lr){6-9} 
 & AUC $\uparrow$  & Precision $\uparrow$ & Recall $\uparrow$ & F1 score $\uparrow$ & AUC $\uparrow$  & Precision $\uparrow$ & Recall $\uparrow$ & F1 score $\uparrow$\\
\hline
0.1 & \textbf{0.9426} & \textbf{0.8511} & \textbf{0.8108} & \textbf{0.7979} & \textbf{0.9022}  & \textbf{0.7707} & \textbf{0.7568} & \textbf{0.7536} \\
1.0 & 0.9258  & 0.6887 & 0.7297 & 0.7066 & 0.8790  & 0.7162 & 0.7297 & 0.7156 \\
10 & 0.9235  & 0.6597 & 0.7027 & 0.6555 & 0.8435  & 0.6283 & 0.6757 & 0.6388 \\

\hline
\end{tabular}
\end{center}
\label{tab:abl_beta}
\end{table*}

\begin{figure*}[]
\centering
\includegraphics[width=1.0\linewidth]{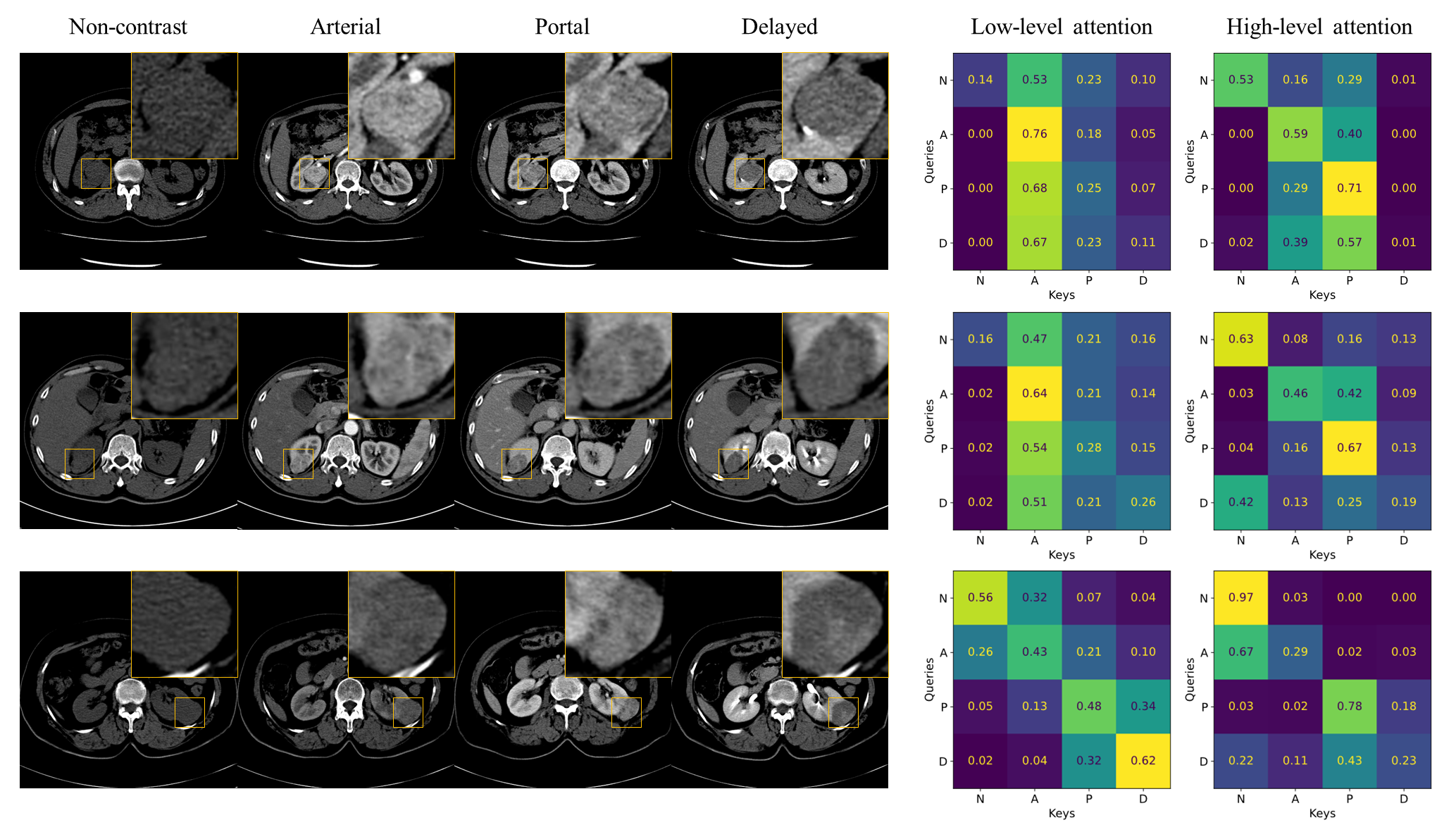}
\caption{Visualization of low-level and high-level inter-phase attention matrices ($A^{low}, A^{high}$) extracted by our LACPANet for cancer subtype classification on test data. 
The first row shows the case of patients with oncocytoma. For this case, the baseline (3D multi-phase baseline) misclassifies the cancer subtype label as ccRCC, while our proposed LACPANet correctly classifies the subtype as oncocytoma.
The second row shows the AML subtype case, and for this case, the baseline misclassifies the cancer label as chRCC, while our model makes the correct classification.
The last row represents the results for the ccRCC case. The baseline wrongly predicts the cancer subtype as chRCC, while our method classifies the cancer subtype correctly from the input CT scan.
In each row, input four-phase CT images, including non-contrast, arterial, portal, and delayed phases, are shown on the left, and the low- and high-level attention matrices obtained from our LACPANet are shown on the right.
We show the representative slices of CT scans for visualization, and the tumor regions are zoomed-in (yellow) on each phase.
The attention value for each query-key pair is displayed on the matrix.
}
\label{fig:attention}
\end{figure*}

\subsection{Hyper-Parameter Ablation Study}\label{sec:ablation}
This section provides detailed ablation studies to investigate the influence of the hyper-parameters included in the proposed method.
We present the performances of our LACPANet by varying $\lambda$, $\alpha$, and $\beta$ under both the semi-automated and fully-automated settings. 
$\lambda$ balances the contributions between the two terms in (\ref{eqa:residual}).
As $\lambda$ increases, the contribution degree of the weighted average vectors of value matrix $AV$ also increases. 
Table~\ref{tab:abl_lambda} shows the results for different values of $\lambda \in \{0.01, 0.1, 1.0\}$, where the performance of our model is evaluated without multi-scale attention (``Ours w/o multi-scale attention'') to first determine $\lambda$ independently from other parameters.
As shown in the table, we find that $\lambda=0.1$ achieves the best performance in all evaluation metrics for both settings.
Therefore, we fix $\lambda$ as 0.1 by default for other experiments.

We also conduct several experiments to investigate the influence of the hyper-parameter $\alpha$, which is used for balancing the trade-off between low-level and high-level predictions, \textit{i.e.}, $\hat{y}^{low}$ and $\hat{y}^{high}$.
A larger $\alpha$ increases the degree of contribution of low-level predictions compared with high-level ones in the final cancer subtype prediction result.
We show the performance of our LACPANet by varying $\alpha$ from 0.1 to 0.9 with a step size of 0.2 in Table~\ref{tab:abl_alpha}.
It can be seen that the best result is obtained when $\alpha$ is 0.7. By default, we set $\alpha=0.7$ in all tests.

Finally, we study the influence of $\beta$, which balances the low-level and high-level classification loss, with different values $\beta \in \{0.1, 1.0, 10\}$. As $\beta$ increases, the contribution degree of the high-level classification loss on the model training also increases. The results are summarized in Table~\ref{tab:abl_beta}.
We find that the best classification performance is achieved with $\beta=0.1$ under both semi-automated and fully-automated settings. Hence, we empirically set $\beta=0.1$ as the default setting.

\subsection{Visualization of Attention Response}\label{sec:visualize_attention}
To further understand how inter-phase lesion-aware attention works, we visualize segmented tumor regions of multiple phases and attention response, \textit{i.e.}, attention matrix $A$ in (\ref{eqa:attention}).
We compare the prediction results of our model with the baseline and present low-level and high-level attention matrices, $A^{low}$ and $A^{high}$, extracted from the LACPANet, which are the core components of our model for performance improvement.     
We demonstrate three patient cases in Fig.~\ref{fig:attention}, where the rows in the figure indicate different patient cases.  
We first note that in these cases, the 3D multi-phase baseline fails to correctly classify cancer subtypes from input CT scans, while our model makes the right prediction.
For example, in the first case (first row), the baseline misclassifies the cancer subtype as ccRCC, where the actual subtype label is oncocytoma.
Differentiating renal oncocytoma from ccRCC is challenging to some degree, even for experienced radiologists, due to overlapping imaging features among them~\cite{WHO_2016}.
Meanwhile, our method correctly classifies the renal mass subtype by utilizing the proposed inter-phase lesion-aware attention scheme which enhances multi-phase lesion features.
We can clearly observe that, in low-level and high-level attention matrices, attention values of the arterial phase to any query phases are high, which means that lesion features of the arterial phase contribute a lot to enhancing the features of other phases to aid the classification of tumor subtypes.
This phenomenon can be partially explained by clinical findings from some studies~\cite{Tanaka_deep, GAKIS2011274} that the arterial phase exhibits the most useful imaging features for differentiating oncocytoma and ccRCC.

In the second case (second row in the figure), the ground-truth tumor subtype label is AML, but the baseline predicts the class label as chRCC.
On radiographic images, chRCC is often confused with AML~\cite{chRCC_AML}, and since AML is benign, whereas chRCC is malignant, incorrect differentiation between them might lead to unnecessary surgery.
In contrast, our LACPANet correctly classifies the tumor of the input CT scan as AML for this case.
From the low-level attention matrix, we can see that three dynamic contrast-enhanced CT phases attend to each other to enhance lesion features across phases, where the arterial phase draws the highest attention values.
This implies that our 3D lesion-aware inter-phase attention can capture inter-dependencies of lesion features across phases for accurate classification of renal tumors on multi-phase CT.

The third case presents the results for ccRCC patient CT data.  
In this case, the baseline mistakenly classifies the tumor subtype into chRCC, while our model accurately classifies the subtype.
From the attention matrices in the figure, we see that neighboring phases for each query phase have high attention values, which implies that the neighboring phases contain informative features to interpret the features of the query phase.
Differential diagnosis of ccRCC and chRCC is critical in making treatment planning since chRCC is often managed by partial nephrectomy, while ccRCC is recommended to be treated with radical resection~\cite{chrcc_behavior, triple-cls}.  
Hence, these results demonstrate the effectiveness of our proposed LACPANet in classifying renal tumor subtypes on multi-phase CT.

\section{Discussion}\label{sec:discussion}
Regarding the dataset, although all patients underwent surgery in one hospital, their CT scans were acquired in various hospitals, as explained in Section~\ref{sec:dataset}, thus our model is believed to have generalizability to CT equipment in various hospitals.
Nevertheless, there are some limitations in this work.
The CT images in the dataset were collected from retrospective cohorts of patients, and only patients who underwent nephrectomy, even for benign tumors, were selected, which may introduce selection bias.
The dataset does not include CT data for other geographic regions, therefore further research on the broader population is needed to investigate potential overfitting and validate the generalizability of our model.
Currently, there is a lack of other four-phase CT datasets for renal tumor subtype classifications.
In this work, we developed a framework that requires complete four-phase CT.
It would be beneficial to expand the model to be applicable for CT scans with fewer than four phases with arbitrary phase combinations.


\section{Conclusion}\label{sec:conclusion}
In this paper, we present LACPANet, a novel lesion-aware cross-phase attention network for kidney cancer diagnosis on multi-phase CT scans. LACPANet is designed to attend different CT phases by leveraging inter-dependencies between lesion-aware features that are captured via the transformer-based attention mechanism.
Also, we introduce a multi-scale attention scheme to exploit cross-phase information at multiple feature levels. Extensive experiments have demonstrated that our model outperformed prior kidney cancer diagnosis models, and we have validated the effectiveness of our core components through ablation studies. In the future, we will consider incomplete CT data with missing phases to make our model more applicable to clinical scenarios.  
\section*{Acknowledgement}
This work was supported by the Korea Medical Device Development Fund grant
funded by the Korea government (the Ministry of Science and ICT, the Ministry of Trade,
Industry and Energy, the Ministry of Health \& Welfare, the Ministry
of Food and Drug Safety) (Project Number: 1711195432, KMDF$\_$PR$\_$20200901$\_$0096).
\printcredits

\bibliographystyle{cas-model2-names.bst}

\bibliography{cas-refs}



\end{document}